# A Mixed-Method Landscape Analysis of SME-focused B2B Platforms in Germany

## Tina Krell*

Humboldt Institute for Internet and Society, 10117 Berlin, Germany.
E-mail: tina.krell@hiig.de

## Dr. Fabian Braesemann

University of Oxford, Saïd Business School, Oxford OX1 1HP, UK.
University of Oxford, Oxford Internet Institute, Oxford OX1 3JS, UK.

## Dr. Fabian Stephany

University of Oxford, Oxford Internet Institute, Oxford OX1 3JS, UK.
Humboldt Institute for Internet and Society, 10117 Berlin, Germany.

## Dr. Nicolas Friederici

Humboldt Institute for Internet and Society, 10117 Berlin, Germany.

## Philip Meier

Humboldt Institute for Internet and Society, 10117 Berlin, Germany.

* Corresponding author

**Abstract:** Digital platforms offer vast potential for increased value creation and innovation, especially through cross-organizational data sharing. It appears that SMEs in Germany are currently hesitant or unable to create their own platforms. To get a holistic overview of the structure of the German SME-focused platform landscape (that is platforms that are led by or targeting SMEs), we applied a mixed method approach of traditional desk research and a quantitative analysis. The study identified large geographical disparity along the borders of the new and old German federal states, and overall fewer platform ventures by SMEs, rather than large companies and startups. Platform ventures for SMEs are more likely set up as partnerships. We indicate that high capital intensity might be a reason for that.

**Keywords:** SME; platforms; data cooperation; digital platforms; Germany; landscape analysis; cluster analysis; mixed-method; governance; Mittelstand

# 1 Introduction

*Overview*

Digital platforms offer vast potential for increased value creation and innovation, especially through cross-organizational data sharing (Parker & Van Alstyne, 2018). It appears that SMEs in Germany are currently hesitant or unable to create their own data platforms. Platform enterprises specifically targeting SMEs are not immediately visible and have not yet been studied rigorously. An emerging practice and policy literature broadly acknowledges problems and opportunities of digital platforms for German SMEs (BMWi, 2019a; Busch, C., 2019). Other work relates primarily to Industry 4.0 or the Internet of Things (IoT) (BMWi, 2019b; Riemensperger & Falk, 2019b) without focusing on cooperation among medium-sized companies. As current literature mainly takes the perspective of existing platform owners or individual users (McIntyre & Srinivasan, 2016), practical cooperative potential of platforms for SMEs cannot yet be conclusively derived. Helfat & Raubitschek (2018) claim that platform owners generally display strong capabilities to innovate, to sense strategic opportunities and risks in the marketplace, and integrative capabilities to handle the complexity of multi-sided business models. This is supported by the observation that platforms driven by large technology companies such as Siemens Mindsphere or young startups like Kreatize are currently the best-recognized data platform offerings for SMEs. Government-led data platform setup projects, such as GaiaX, have recently been advanced but their feasibility is still to be verified. The emergence or failure of new platforms is generally an understudied subject (Gawer & Cusumano, 2013).

*Defining the problem*

There lies much potential in studying viable models of data-based platform cooperation in Germany. To do this holistically, we need a comprehensive overview of the German SME-focused platform landscape, that is platforms led by or targeting SMEs. We therefore designed a mixed-method study, through traditional desk research and quantitative analysis based on data from a globally leading platform which provides information on private and public companies, with a focus on early-stage firms. Through the desk research we captured the German SME platform landscape on the micro-level and compared the companies on different dimensions. The quantitative analysis contrasts and elevates the findings of the desk research, by providing a large-scale overview of the German digital landscape and the centrality of SME-oriented platform firms within that universe. On the micro-level we focused on the organisational structure. To provide the large-scale overview, we focus on the three production factors that are combined by firms in the knowledge economy: technologies, financial capital, and human capital.

Both methods were conducted independently and brought together to answer the following questions:

RQ1: What is the structure of the German platform economy landscape and how are SME-focused firms integrated?

    RQ1a: Who sets up German SME-focused platforms and how are they operated?

    RQ1b: What are major obstacles for SME-focused platforms?

RQ1c: Which technological and business models are predominant (technology perspective)?

RQ1d: Is there a common financial backbone of venture capital and public funders providing capital to SME-focused platforms (capital perspective)?

RQ1e: Are SME-focused platforms integrated in the knowledge flows of the German innovation ecosystem associated with entrepreneurial mobility (human capital perspective)?

*Structure of the paper*

In the following section, we describe data, methods and preliminary findings of the desk research. In the third section we describe data, method and preliminary findings of the quantitative analysis. Key findings are consolidated in the fourth section, which also provides an outlook.

## 2 Desk Research

*Data and Methods*

We define German SMEs along the EU convention (European Commission, 2003) by staff headcount, turnover or balance sheet total. Additionally, we expand the search by the definition of "Mittelstand" companies. That is, companies that hold up to two natural persons or their family members (directly or indirectly) at least 50% of the shares in a company, these natural persons belong to the management (IfM Bonn, n.a.). We collected platforms defined by Gawer's (2014) integrative framework, as double-sided markets with a modular technological architecture. To be collected, a platform needed to fulfill at least one of the criteria.

Our focus was on platforms in the B2B sector in Germany, which are an understudied subject. Prioritising literature with that focus meant that we mostly observed grey literature from governmental institutions, reports, articles and white paper from interest groups, research organisations and consulting firms. If applicable, we incorporated academic literature and news articles, and collected companies from there. (A comprehensive list can be shared on request.)

Additionally, we examined platform-relevant conferences and events for German SME-focused platforms. Conferences that were considered relevant, needed to fulfill the criteria to be established in the B2B sector and, or in the platform community in Germany and Europe, such as the NOAH conference, the Platform Economy Summit and the Hannover Messe.

For comparison, identified companies were collectively broadened by dimensions and standardised attributes such as regional base, company size, ownership structure, industry and sector.

*Preliminary Findings*

The two-fold approach gave us the non-exhausted list of 137+ SME-focused platforms, and a more granular understanding of the dimensions described above.

Overall, platform operators are largely made up of startups and large companies, medium-sized companies are less to be found in the operator role. Also, all in all, platforms are based in large cities.

We found the largest concentration of platforms by far in Berlin, most of them being startups. Upon closer examination, startups were partially set up independently with own technology and investment strategy, but also as partnerships with, or made by large companies from South Germany. The platforms in North Rhine-Westphalia, Baden-Württemberg and Bavaria were owned, made and partnered by established large companies, that identified as family owned "Mittelstand" companies.

On closer examination, often headquarters of the parent company and the seat of the platform are not identical, here we differentiated whether the platform was still in the same city (CheMondis instead of Messe Deutz city center) or even had offices opened in other cities (XOM Materials from Klöckner, which have its headquarters in Duisburg, platform is headquartered in Berlin).

Remarkable findings to us were stark regional differences. We identified most SME-focused platforms in Berlin and Germany's old federal states, such as North Rhine-Westphalia, Bavaria and Baden-Württemberg. Even after purposefully searching, we could identify only one B2B platform in one of the new federal states, Saxony. Prisma Capacity, a European online platform for trading capacity rights in the oil and gas sector. Another Saxony company we identified is the evan.network, an independent non-commercial startup, running an open (access) blockchain ecosystem, compliant with GDPR.

Evenly remarkable is the insight into the founding structure of the identified platform companies. By far, most platforms are genuinely built by large companies and startups, roughly a fifth of the identified platform ventures are partnerships. If Mittelstand companies build platforms they do it more likely as a partnership, rather than building one themselves. If large companies partner up for a platform, they likely do this with other evenly large and established companies.

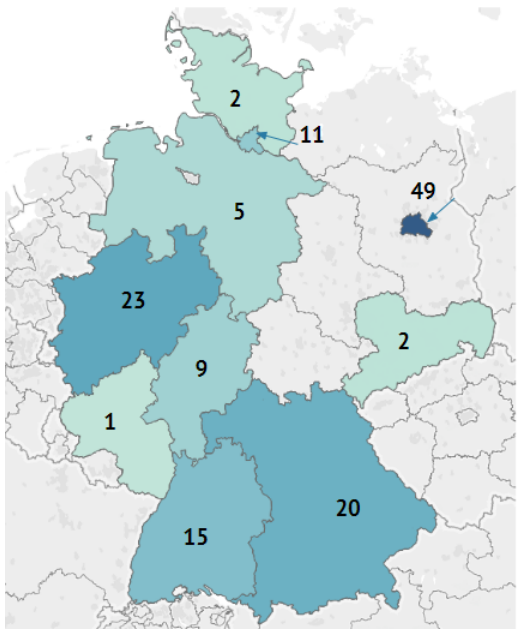

**Figure 1** Geographic overview and count of Platform headquarters by federal state.

## 3 Quantitative Analysis

*Data and Methods*

Our data are extracted from Crunchbase, a global database of startup companies. Crunchbase sources its data in various ways: a partner program where venture capital companies provide information on their portfolio companies in exchange for data access and crowdsourced data from companies verified by in-house data team. While Crunchbase data is not representative of the German economy as a whole it has a wide coverage among technology companies and digital platforms. Thus, it can be argued that firms covered by the Crunchbase data represent the domestic innovation ecosystem, in which the SME-focused platforms are embedded. Crunchbase provides detailed information about the firms listed on the website. In particular, it lists investments and funding information, the entrepreneurs and leading managers of the firms and a brief description of companies' business models and tags describing the technologies used. Crunchbase currently lists more than 13,000 firms in the German innovation ecosystem.

In order to obtain the networks of the German innovation ecosystem, we have processed the data in multiple steps. First, to construct the network of technologies in the German innovation ecosystem, we transform the list of technology-tags of all 13,000 German firms into an adjacency list based on a binary vector of tech-labels for each firm (Braesemann, 2019; Braesemann & Baum, 2020). In the resulting network, two firms are connected if they share at least one technology label. This complex network is characterised by dense groups of firms and technologies forming natural clusters which can be identified by unsupervised statistical learning. We apply the Louvain method for cluster detection based on network modularity to identify the clusters.

In order to reveal the financial backbone, we construct a bipartite network of investors and firms from all companies in the dataset. Here, a connection is established between a firm and an investor if the firm obtained funding from the investor. To assess the know-how transfer between firms in the German landscape, we construct a third network between firms based on the mobility of entrepreneurs: e.g. two firms could be connected because the CTO from firm A became senior manager at firm B, or because CEO of firm C founded another company D.

In all these networks, we investigate the integration of the SME-oriented platforms that were identified by desk research and we compare vectors of (a) technologies, (b) investors, (c) entrepreneurs to find the closest matching firms within the German firm ecosystem. The SME-oriented platforms, their closest 'cousins' identified by the matching procedure and a random sample of non SME-oriented firms in the dataset are then compared to reveal the extent to which platforms focusing on SMEs are relevant within the overall German digital economy.

*Preliminary Findings*

The findings of the landscape analysis are summarised in figure 1 to figure 6. We have considered a network perspective for each of the three production factors: technologies, financial capital, and entrepreneurial know-how. Figure 1 shows the network of technologies. It displays a bipartite network with the firms and technologies as small nodes. Additionally, we have highlighted the SME-oriented platform firms. The data naturally groups into seven clusters (Louvain Method). These are highlighted by different colours. As can be seen from both the network and the barchart, most SME-oriented platform firms tend to be central parts of the technology network in three clusters: IT, E-Commerce, and Manufacturing. Most other parts of the network are home to less SME-oriented platforms. They are almost absent in the clusters Healthcare and Consumer. Given the small size of the Education cluster, SME-oriented platforms are relatively frequent.

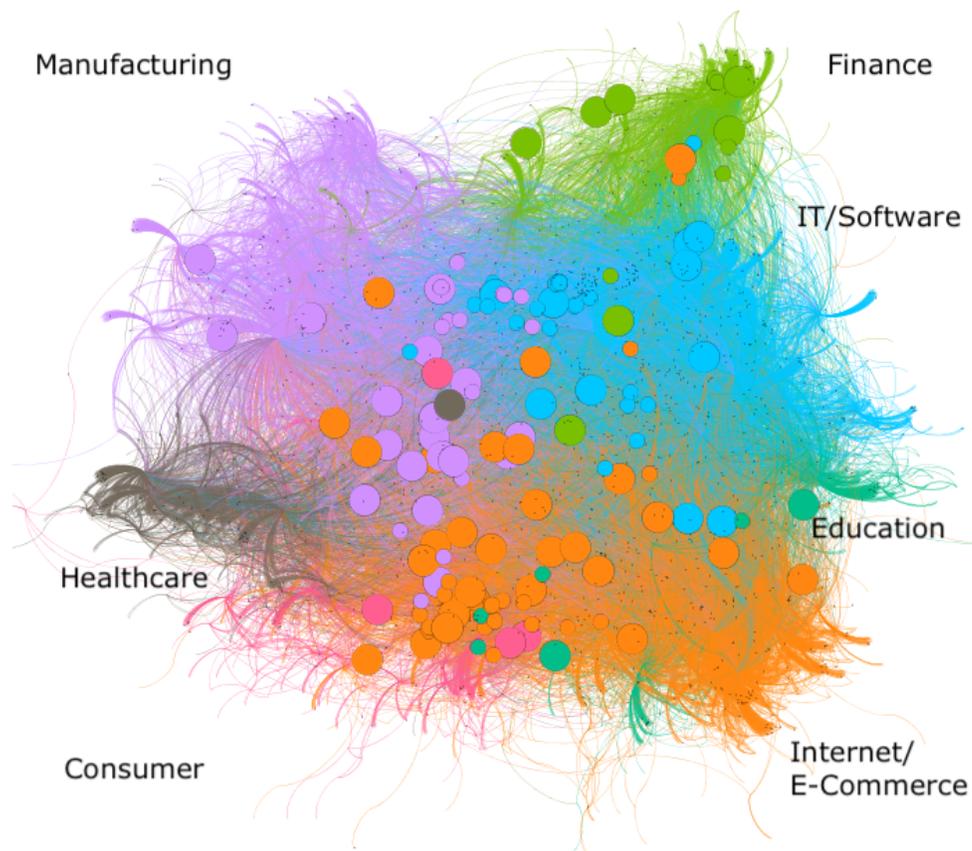

**Figure 2** Technology network of the German innovation ecosystem. The data naturally clusters in seven group (colours). Most of these groups are characterised by dense interactions within and between the groups. Most of the SME-oriented platforms (large nodes) are relatively central in the network, with most firms being part of the groups IT/Software, Internet / E-Commerce, and Manufacturing.

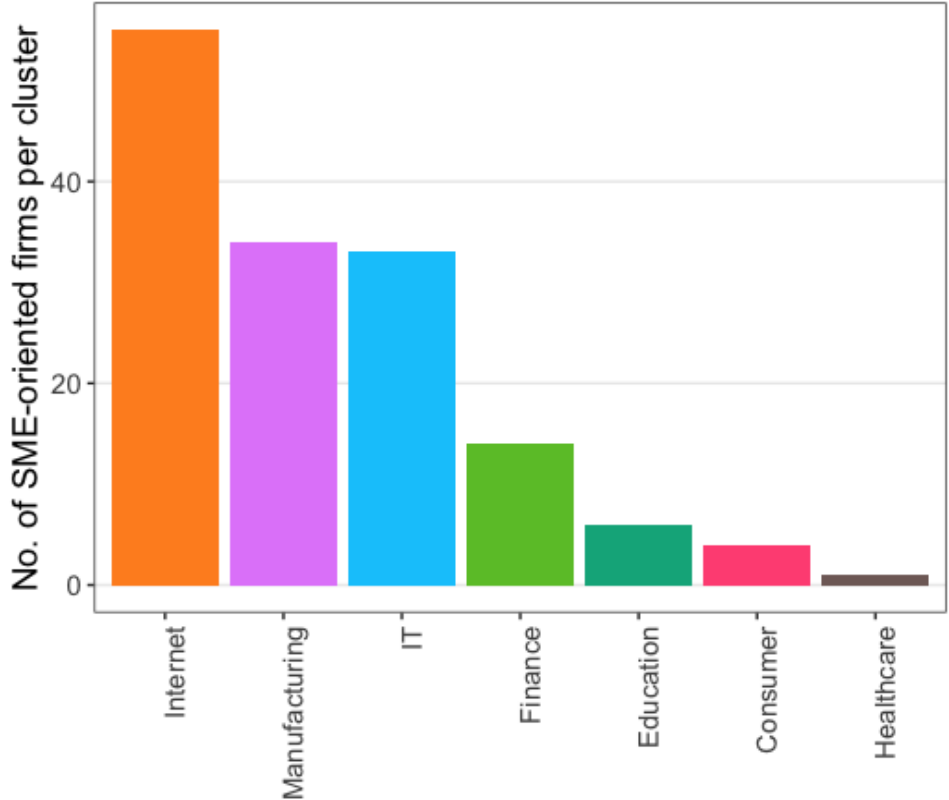

**Figure 3** Number of SME-oriented platforms per technology group. The majority of firms clearly cluster in three groups: Internet / E-Commerce, Manufacturing, and IT. In comparison, the other parts of the German digital economy have fewer SME-oriented platforms.

Particularly relevant for the growth oriented innovation ecosystem is the availability of venture capital funding and the ability of firms to interact with different potential funders to obtain a solid funding base and to mitigate the risk of dependence from individual capitalists. The overall funding landscape is displayed in figure 3. It shows the firms as nodes connected by flows of capital if two firms had the same funder. It can be seen from the network, that most firms are connected only in smaller clusters, long links connecting different parts of the network are rare. Nonetheless, there is a core component of firms that are connected by many links. This core is home to the some of SME-oriented platform firms. Consequently, as displayed in figure 4, the SME-oriented platforms were able to attract higher than average funding (left panel). However, the overall centrality of SME-oriented platforms is not higher than average. This finding could be interpreted as a sign of relative funding-diversity. Funding seems to come from different sources and there is no tendency to support a core group of SME-oriented firms.

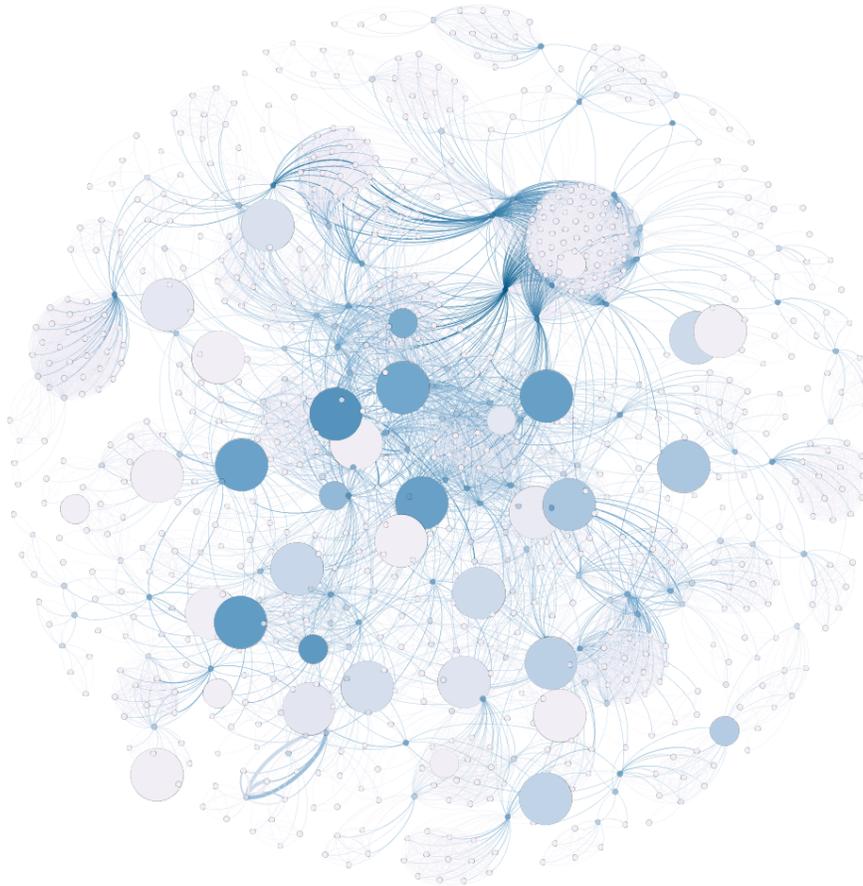

**Figure 4** Funding-Network (colour of nodes - betweenness centrality, darker nodes represent more funding diversity and connections, larger nodes represent SME-oriented firms): The overall network seems to be characterised by many small clusters; nonetheless, some firms form a core component, which also includes some of the SME-oriented platforms. In general, financial capital seems to come from different sources and does not seem to be oriented towards one central group of firms.

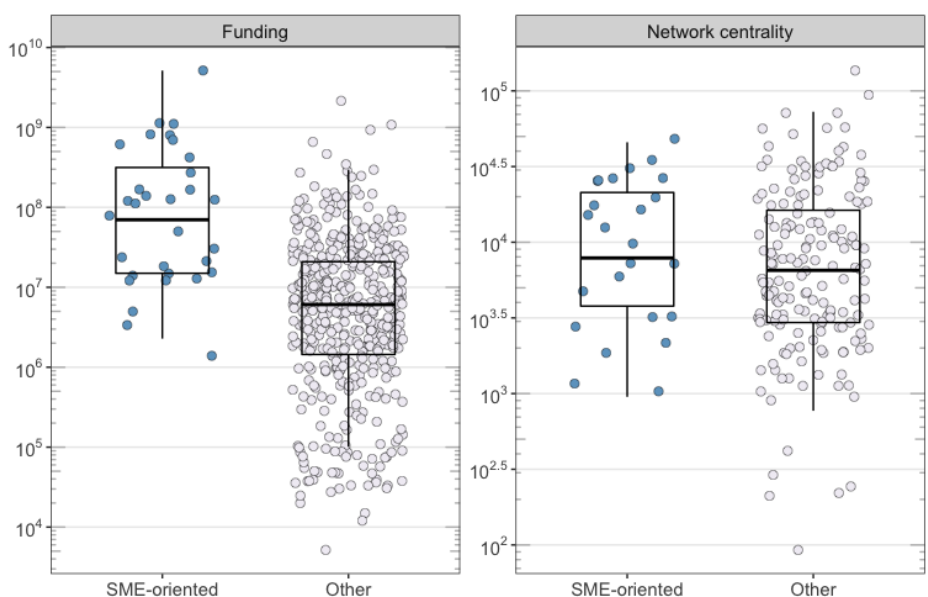

**Figure 5** Funding (left panel) and network centrality within the funding network (right panel) of SME-oriented platform firms and other companies. While SME-oriented platforms tend to be better equipped in terms of venture capital funding, their centrality does not differ substantially from the centrality of other firms in the network.

Besides financial capital, human capital, that is the know-how and experience of entrepreneurs and leading employees are important ingredients for the success of early ventures in the digital sphere. Accordingly, we have displayed the interactions of firms in the German digital economy as a network in figure 5. Firms are connected, if an entrepreneur or employee has been active in both firms. The colour of the nodes corresponds to their betweenness centrality and lager nodes represent SME-oriented platforms.

In contrast to the financial network, SME-oriented platforms tend to cluster substantially in the central component of the network. They are involved in numerous know-how transfers via interactions with other companies in the German innovation ecosystem. This is highlighted in figure 6, which shows the number of employees (panel A) and the network centrality (panel B). SME-oriented platforms seem to be larger firms than average and their centrality tends to be higher than that of the average firm.

We summarise that SME-oriented platform firms are not a peripheral phenomenon. In contrast, such firms are densely connected with other firms in the German landscape, they tend to be solidly equipped with financial capital and they have connections with know-how flows in the system. However, such business models seem to be focused mostly in the three central technological domains of the German innovation ecosystem: in E-commerce, in IT, and in Manufacturing. Other elements of the German digital economy might benefit from more SME-oriented platforms.

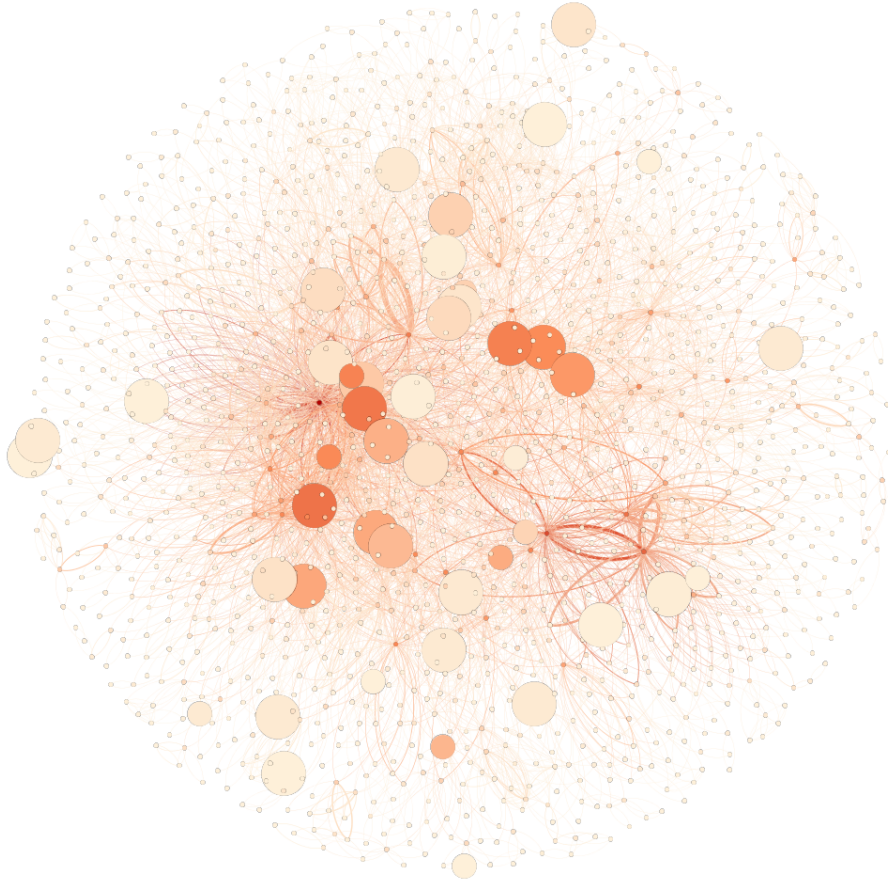

**Figure 6** Network of entrepreneur (colour of nodes - betweenness centrality, dark nodes represent more connections with other companies in terms of employee transfer, larger nodes represent SME-oriented firms): A dense core of the network is shaped by numerous interactions between companies, indicating a vivid exchange of know-how through entrepreneur / employee migration; many of the SME-oriented platform companies are part of the dense core.

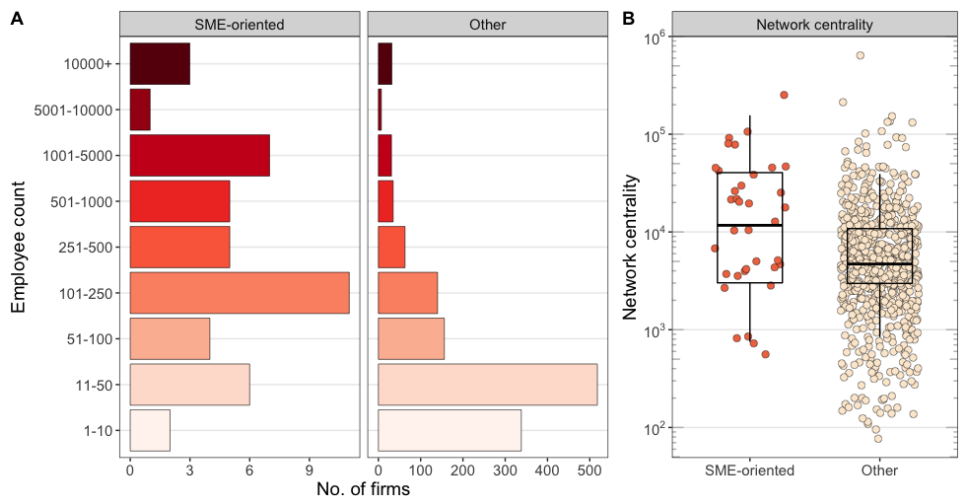

**Figure 7** (A) No. of employees in SME-oriented firms and other companies, (B) Network centrality of firms in entrepreneur-network. The SME-oriented platform companies tend to be larger in terms of employee size and more central in the entrepreneur network.

## 4 Consolidated Findings

As we have seen, there is a large geographic disparity of platform ventures that have their headquarters almost exclusively in the old German federal states or the German capital, Berlin. Platforms are mostly genuinely built by large companies or startups and less so in partnerships. If we find platform ventures as partnerships though, then it is Mittelstand companies that tend to take more advantage of that.

SME platforms have large capital investments, attracting more capital than the median German tech companies. They seem to be well imbedded in the German tech landscape. They are capable of attracting both capital and technological know how in the form of executives and employees and compete for such resources with other German tech companies.

Capital intensity seems to be a strong indicator why most platform ventures are set up by large companies and startups, rather then Mittelstand companies.

---------------------------------------------------------------------------------------------------------